\documentclass[aps,prl,nofootinbib,twocolumn,preprintnumbers,floatfix,superscriptaddress,longbibliography]{revtex4-1}
\usepackage[colorlinks, linkcolor=red, anchorcolor=blue, citecolor=blue, colorlinks=true, urlcolor=blue]{hyperref}
\usepackage{amsmath}
\usepackage{graphicx}
\usepackage{indentfirst}
\usepackage{subfigure}
\usepackage[figuresright]{rotating}
\usepackage{amssymb}
\usepackage{gensymb}
\usepackage{multirow}
\usepackage{fontawesome}
\usepackage{ulem}
\usepackage{bm}
\usepackage{xspace}
\usepackage{threeparttable}
\usepackage{amsthm}
\usepackage{mathrsfs}
\usepackage{epstopdf}
\usepackage{breakurl}
\usepackage{lipsum}
\begin{document}

\title{Light QCD Axion Dark Matter from Double Level Crossings}

\author{Hai-Jun Li} 
\email{lihaijun@itp.ac.cn}
\affiliation{Key Laboratory of Theoretical Physics, Institute of Theoretical Physics, Chinese Academy of Sciences, Beijing 100190, China}
\affiliation{Center for Advanced Quantum Studies, Department of Physics, Beijing Normal University, Beijing 100875, China}

\author{Ying-Quan Peng}
\email{yqpenghep@mail.bnu.edu.cn}
\affiliation{Center for Advanced Quantum Studies, Department of Physics, Beijing Normal University, Beijing 100875, China}

\author{Wei Chao} 
\email{chaowei@bnu.edu.cn}
\affiliation{Center for Advanced Quantum Studies, Department of Physics, Beijing Normal University, Beijing 100875, China}

\author{Yu-Feng Zhou}
\email{yfzhou@itp.ac.cn}
\affiliation{Key Laboratory of Theoretical Physics, Institute of Theoretical Physics, Chinese Academy of Sciences, Beijing 100190, China}
\affiliation{School of Physical Sciences, University of Chinese Academy of Sciences, Beijing 100049, China}
\affiliation{School of Fundamental Physics and Mathematical Sciences, Hangzhou Institute for Advanced Study, UCAS, Hangzhou 310024, China}
\affiliation{International Centre for Theoretical Physics Asia-Pacific, Beijing/Hangzhou, China}

\preprint{ITP-23-205, BNU-23-121}

\date{\today}

\begin{abstract}

The even light QCD axion called the $Z_{\mathcal N}$ axion can both solve the strong CP problem and account for the dark matter (DM).
We point out that the single and double level crossings can naturally take place in the mass mixing between the $Z_{\mathcal N}$ axion and axionlike particle (ALP).
The first level crossing occurs much earlier than the QCD phase transition, while the second level crossing occurs exactly during the QCD phase transition if it exists.
We also find that the single level crossing can transform into the double level crossings, depending on the ALP mass $m_A$ versus the zero-temperature $Z_{\mathcal N}$ axion mass $m_{a,0}$.
Compared with the no level crossing case, the $Z_{\mathcal N}$ axion relic density can be suppressed in the single level crossing, and enhanced or suppressed in the double level crossings.

%\pacs{ }
%\keywords{ }

\end{abstract}
\maketitle

%\section{Introduction}
%{\bf Introduction.}---%

{\it Introduction.}
The QCD axion was predicted by the Peccei-Quinn (PQ) mechanism \cite{Peccei:1977ur, Peccei:1977hh} with a spontaneously broken global $\rm U(1)_{PQ}$ symmetry  to solve the strong CP problem in the Standard Model (SM) \cite{Weinberg:1977ma, Wilczek:1977pj, Kim:1979if, Shifman:1979if, Dine:1981rt, Zhitnitsky:1980tq}.
It is a light pseudo-Nambu-Goldstone boson, acquiring a tiny mass from the QCD non-perturbative effects \cite{tHooft:1976rip, tHooft:1976snw}.
When the QCD instanton generates the QCD axion potential, the axion will settle down at the CP conservation minimum value, solving the strong CP problem.
The QCD axion is also the leading dark matter (DM) candidate, which can be non-thermally produced in the early Universe through the misalignment mechanism \cite{Preskill:1982cy, Abbott:1982af, Dine:1982ah}.
In this case, the axion is massless at high cosmic temperatures and obtains a non-zero mass at the QCD phase transition, then it starts to oscillate when this mass is comparable to the Hubble parameter, explaining the observed cold DM abundance.
 
The reduced-mass QCD axion, the $Z_{\mathcal N}$ axion \cite{Hook:2018jle}, can also solve the strong CP problem \cite{DiLuzio:2021pxd} and account for the DM \cite{DiLuzio:2021gos}.
In this case, the $\mathcal N$ mirror and degenerate worlds that are nonlinearly realized by the axion field under a $Z_{\mathcal N}$ symmetry can coexist in Nature, one of which is the SM world.
In order to solve the strong CP problem, here $\mathcal N$ should be odd and $\mathcal N\geqslant3$.
Since the non-perturbative effects on axion potential from the $\mathcal N$ degenerate QCD groups are suppressed, the resulting $Z_{\mathcal N}$ axion mass is exponentially lighter than the canonical QCD axion, which is independent of the details of the putative UV completion \cite{DiLuzio:2021pxd}.
Additionally, the $Z_{\mathcal N}$ axion can account for the DM through trapped misalignment mechanism \cite{DiLuzio:2021gos}, which is also a natural source of the kinetic misalignment mechanism \cite{Co:2019jts}. 

The axionlike particle (ALP), predicted by several extensions of the SM \cite{Green:1984sg, Witten:1984dg}, is also the DM candidate \cite{Cadamuro:2011fd, Arias:2012az, Chao:2022blc} but not have to solve the strong CP problem.
Considering the canonical QCD axion interacting with the ALP, the cosmological evolution called the level crossing can take place if there is a non-zero mass mixing between them \cite{Hill:1988bu, Kitajima:2014xla, Daido:2015cba, Daido:2015bva, Ho:2018qur, Gavela:2023tzu, Murai:2023xjn, Cyncynates:2023esj, Li:2023xkn}. 
The general condition for level crossing to occur is that the mass and decay constant of the ALP are both much smaller than that of the QCD axion, which can induce the adiabatic transition of these axions.
This adiabatic transition is similar to the MSW effect in neutrino oscillations, leading to the suppression of the axion energy density and isocurvature perturbations.

In this letter, we investigate the $Z_{\mathcal N}$ axion interacting with the ALP, and point out that the single and double level crossings can take place in their mass mixing.
The first level crossing occurs much earlier than the QCD phase transition, while the second level crossing occurs exactly during the QCD phase transition.
The cosmological evolution of axions will become significantly different due to the presence of the second level crossing.
We also estimate the current $Z_{\mathcal N}$ axion relic density through the trapped+kinetic misalignment mechanism.
Compared with the no level crossing case, the $Z_{\mathcal N}$ axion relic density can be significantly modified in the single and double level crossings.

{\it Framework.} 
We consider the $Z_{\mathcal N}$ axion $\phi$ interacting with the ALP $\psi$ through following low-energy effective Lagrangian
\begin{eqnarray}
\begin{aligned}
\mathcal{L}&\supset\frac{1}{2}\partial^\mu\phi\partial_\mu\phi +\frac{1}{2}\partial^\mu\psi\partial_\mu\psi\\
&-\frac{m_a^2(T) f_a^2}{{\mathcal N}^2}\left[1-\cos\left({\mathcal N}\frac{\phi}{f_a}\right)\right]\\
&-m_A^2 f_A^2\left[1-\cos\left({\mathcal N}\frac{\phi}{f_a}+\frac{\psi}{f_A}\right)\right]\, ,
\end{aligned}
\end{eqnarray}
where $m_a(T)$ is the temperature-dependent $Z_{\mathcal N}$ axion mass, $m_A$ is the ALP mass, $f_a$ and $f_A$ are the $Z_{\mathcal N}$ axion and ALP decay constants, respectively.
In the large $\mathcal N$ limit, $m_a(T)$ can be approximated by \cite{DiLuzio:2021pxd, DiLuzio:2021gos}
\begin{eqnarray}
m_a(T)\simeq
\begin{cases}
\dfrac{m_\pi f_\pi}{\sqrt[4]{\pi} f_a}\sqrt[4]{\dfrac{1-z}{1+z}}{\mathcal N}^{3/4}z^{\mathcal N/2}\, , &{\rm LT}\\
\dfrac{m_\pi f_\pi}{f_a}\sqrt{\dfrac{z}{1-z^2}}\, , &{\rm MT}\\
\dfrac{m_\pi f_\pi}{f_a}\sqrt{\dfrac{z}{1-z^2}}\left(\dfrac{\gamma T}{T_{\rm QCD}}\right)^{-n}\, , &{\rm HT}
\end{cases} 
\label{maNT}
\end{eqnarray}
where $m_\pi$ and $f_\pi$ are the mass and decay constant of the pion, respectively, $z\equiv m_u/m_d\simeq0.48$, $m_u$ and $m_d$ are the up and down quark masses, $\gamma\in(0,1)$ is a temperature parameter and we set $\gamma=0.1$, $T_{\rm QCD}\simeq150\, \rm MeV$ is the critical temperature of the QCD phase transition, and $n\simeq4.08$.
The LT, MT, and HT represent the low temperatures $T\leq T_{\rm QCD}$, the medium temperatures $T_{\rm QCD} < T \leq T_{\rm QCD}/\gamma$, and the high temperatures $T>T_{\rm QCD}/\gamma$, respectively.
Note that LT corresponds to the zero-temperature $Z_{\mathcal N}$ axion mass $m_{a,0}$, which is suddenly exponentially suppressed ($\propto{\mathcal N}^{3/4}z^{\mathcal N/2}$) at $T_{\rm QCD}$ due to the $Z_{\mathcal N}$ symmetry.
The equations of motion (EOM) of $\phi$ and $\psi$ are
\begin{eqnarray}
\begin{aligned}
\ddot\phi&+3H\dot\phi+\frac{m_a^2(T) f_a}{\mathcal N}\sin{\left({\mathcal N}\frac{\phi}{f_a}\right)}\\
&+{\mathcal N}m_A^2 \dfrac{f_A^2}{f_a} \sin\left({\mathcal N}\frac{\phi}{f_a}+\frac{\psi}{f_A}\right)=0\, ,
\end{aligned}
\end{eqnarray}
\begin{eqnarray}
\ddot\psi+3H\dot\psi+m_A^2 f_A\sin\left({\mathcal N}\frac{\phi}{f_a}+\frac{\psi}{f_A}\right)=0 \, , 
\end{eqnarray}
where the dots represent derivatives with respect to the physical time $t$, and $H(T)$ is the Hubble parameter.
Considering the oscillation amplitudes of the axion fields are much smaller than the corresponding decay constants, the mass mixing matrix is given by
\begin{eqnarray}
\mathbf{M}^2=
\left(
\begin{array}{cc}
m_a^2(T)+{\mathcal N}^2 m_A^2\dfrac{f_A^2}{f_a^2}  & \quad {\mathcal N}m_A^2\dfrac{f_A}{f_a} \\
{\mathcal N}m_A^2\dfrac{f_A}{f_a} & \quad   m_A^2
\end{array}
\right)\, .
\end{eqnarray}
Then we diagonalize the mass mixing matrix and derive the heavy ($a_h$) and light ($a_l$) mass eigenstates
\begin{eqnarray}
\left(
\begin{array}{c}
a_h \\
a_l 
\end{array}
\right)
=
\left(
\begin{array}{cc}
\cos \alpha & \quad \sin \alpha \\
-\sin \alpha  & \quad   \cos \alpha
\end{array}
\right)
\left(
\begin{array}{c}
\phi \\
\psi 
\end{array}
\right)\, ,
\end{eqnarray} 
with the corresponding mass eigenvalues $m_{h,l}(T)$
\begin{eqnarray}
\begin{aligned}
m_{h,l}^2(T)&=\frac{1}{2}\left[m_a^2(T)+m_A^2+{\mathcal N}^2m_A^2 \dfrac{f_A^2}{f_a^2}\right]\\
&\pm\frac{1}{2f_a^2}\bigg[-4 m_a^2(T) m_A^2 f_a^4\\
&+\left(\left(m_a^2(T)+m_A^2\right)f_a^2+{\mathcal N}^2m_A^2 f_A^2\right)^2\bigg]^{1/2}\, ,~
\end{aligned}
\end{eqnarray} 
where $\alpha$ is the mass mixing angle. 

\begin{figure*}[t]%%%%%%%%%%%%%%%%%%fig_dlc
\centering
\includegraphics[width=0.49\textwidth]{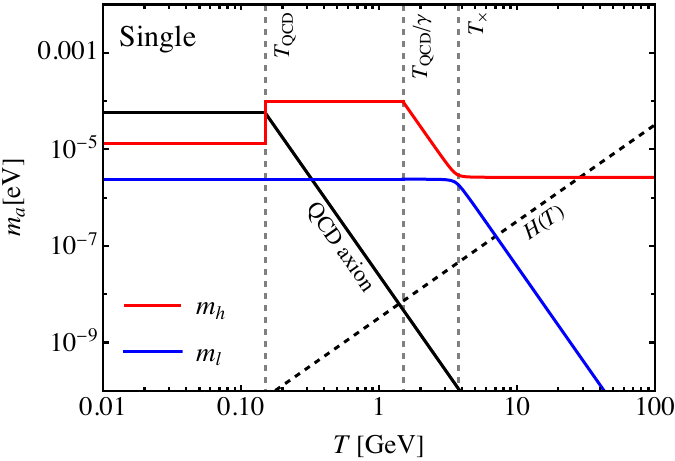}\quad\includegraphics[width=0.49\textwidth]{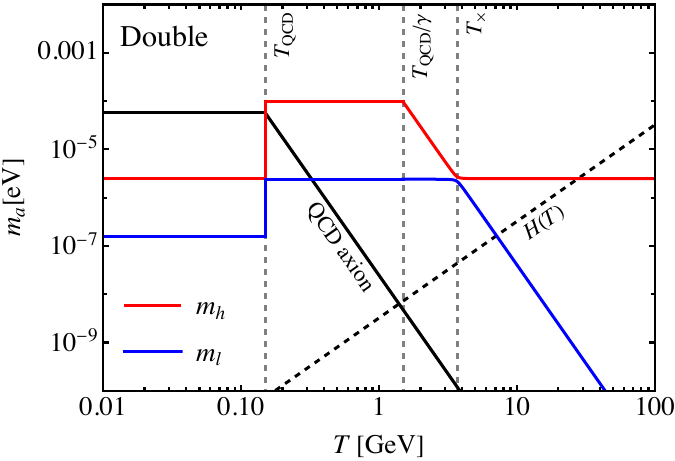}  
\caption{Toy examples of the single level crossing (left) and double level crossings (right).
The red and blue lines represent the temperature-dependent mass eigenvalues $m_h(T)$ and $m_l(T)$, respectively.
The black solid line corresponds to the canonical QCD axion.
The black dashed line represents the Hubble parameter $H(T)$.
The three gray dashed lines (from left to right) represent the temperatures $T_{\rm QCD}$, $T_{\rm QCD}/\gamma$, and $T_\times$, respectively.
Left: we set $f_a=10^{11}\,{\rm GeV}$, $r_m=0.025$ (implies $\Rightarrow \mathcal N \le 13$), $\mathcal N=9$ (implies $\Rightarrow r_f<0.111$), and $r_f=0.05$.
Right: we set $f_a=10^{11}\,{\rm GeV}$, $r_m=0.025$ (implies $\Rightarrow \mathcal N \ge 15$), $\mathcal N=23$ (implies $\Rightarrow r_f<0.043$), and $r_f=0.01$.
}
\label{fig_dlc}
\end{figure*}

{\it Single level crossing.}
Generally, the level crossing can take place when the difference of $m_h^2(T)-m_l^2(T)$ gets a minimum value
\begin{eqnarray}
\frac{{\rm d}\left(m_h^2(T)-m_l^2(T)\right)}{{\rm d}T}\bigg|_{T_\times}=0\, ,
\end{eqnarray} 
where $T_\times$ is the level crossing temperature
\begin{eqnarray}
T_\times = \frac{T_{\rm QCD}}{\gamma}\left(m_{a,\pi}\right)^{1/n}\left(m_A^2-{\mathcal N}^2 m_A^2\dfrac{f_A^2}{f_a^2}\right)^{-1/(2n)}\, ,~ 
\end{eqnarray} 
where we have defined
\begin{eqnarray}
m_{a,\pi}=\dfrac{m_\pi f_\pi}{f_a}\sqrt{\dfrac{z}{1-z^2}}\, .
\end{eqnarray}
Note that this is the {\it first} level crossing in the mass mixing between the $Z_{\mathcal N}$ axion and ALP, in which the $Z_{\mathcal N}$ axion mass is temperature-dependent at $T>T_{\rm QCD}/\gamma$.
In this case, the $Z_{\mathcal N}$ axion mass at $T_\times$ is given by
\begin{eqnarray}
m_a(T_\times)\simeq\left(m_A^2-{\mathcal N}^2 m_A^2\dfrac{f_A^2}{f_a^2}\right)^{1/2}\, ,
\end{eqnarray} 
which should satisfy $0<m_a(T_\times)/m_{a,\pi}<1$.
Then we derive the condition for the first level crossing
\begin{eqnarray}
0<r_m^2-{\mathcal N}^2r_m^2 r_f^2 < 1\, ,
\label{clc_1}
\end{eqnarray} 
where we have defined $r_m=m_A/m_{a,\pi}<1$ and $r_f=f_A/f_a$.
Eq.~(\ref{clc_1}) shows the allowed regions for level crossing with the parameters $r_m$, $r_f$, and $\mathcal N$.
Note that $r_f<1/{\mathcal N}$ and we are more concerned about the case that $r_m$ and $r_f$ are much smaller than 1.
We show the temperature-dependent mass eigenvalues $m_{h,l}(T)$ as a function of the cosmic temperature $T$ in Fig.~\ref{fig_dlc} (left), in which the parameters are selected to satisfy the condition Eq.~(\ref{clc_1}), $\rm i.e.$, ({\it single}) level crossing occurs in this plot.
Here we have set $f_a=10^{11}\,{\rm GeV}$, $r_m=0.025$, $r_f=0.05$, and $\mathcal N=9$.
The red and blue lines represent $m_h(T)$ and $m_l(T)$, respectively.
At high temperatures, the heavy mass eigenvalue $m_h(T)$ corresponds to the ALP, while the light one $m_l(T)$ corresponds to the $Z_{\mathcal N}$ axion.
The $m_h(T)$ and $m_l(T)$ would approach to each other at the level crossing temperature $T_\times$ and then move away from each other.
When $T\leq T_\times$, the $m_h(T)$ represents the $Z_{\mathcal N}$ axion, and the $m_l(T)$ represents the ALP.
In this case, the cosmological evolution of $Z_{\mathcal N}$ axion mass is given by
\begin{eqnarray}
\begin{aligned}
m_a(T)\simeq
\begin{cases}
m_h(T)\, , &T\leq T_\times \\
m_l(T)\, . &T> T_\times
\end{cases} 
\end{aligned}
\end{eqnarray} 
Note that the ALP mass is constant from high to low temperatures.
The above discussion is focused on the ``{\it single level crossing}" case.

%%%%%%%%%%%%%%%%%%%%%%%%%%

{\it Double level crossings.} 
Furthermore, we find an interesting phenomenon that another level crossing can occur during the QCD phase transition for some parameters, see also Fig.~\ref{fig_dlc} (right), resulting in first and second level crossings at $T_\times$ and $T_{\rm QCD}$, respectively. 
Here we have set $f_a=10^{11}\,{\rm GeV}$, $r_m=0.025$, $r_f=0.01$, and $\mathcal N=23$.
In this case, the temperature-dependent behavior of the axions at $T>T_{\rm QCD}$ is similar to the single level crossing case.
However, once the temperature becomes $T\leq T_{\rm QCD}$, the $m_h(T)$ again comprises the ALP and the $m_l(T)$ comprises the $Z_{\mathcal N}$ axion, which is completely different from the single level crossing case and called the ``{\it double level crossings}".
In the double level crossings, the evolution of axion mass is given by
\begin{eqnarray}
\begin{aligned}
m_a(T)\simeq
\begin{cases}
m_l(T)\, , &T\leq T_{\rm QCD} \\
m_h(T)\, , &T_{\rm QCD}<T\leq T_\times  \\
m_l(T)\, . &T>T_\times
\end{cases} 
\end{aligned}
\end{eqnarray}
Compared with the no level crossing case, we find that the $Z_{\mathcal N}$ axion and ALP masses at both high ($T>T_\times$) and low ($T\leq T_{\rm QCD}$) temperatures are unchanged if neglecting the evolutions at $T_{\rm QCD}<T\leq T_\times$.
In order for this double level crossings to occur, we note that the ALP mass should be larger than the zero-temperature $Z_{\mathcal N}$ axion mass $m_{a,0}$, $\rm i.e.$, 
\begin{eqnarray}
m_A>\dfrac{m_\pi f_\pi}{\sqrt[4]{\pi} f_a}\sqrt[4]{\dfrac{1-z}{1+z}}{\mathcal N}^{3/4}z^{\mathcal N/2}\, .
\end{eqnarray}
We also find that if the first level crossing does not occur at $T_\times$, then the second level crossing is also unlikely to take place, $\rm i.e.$, Eq.~(\ref{clc_1}) should also be satisfied.
Therefore, the condition for the double level crossings is
\begin{eqnarray}
\begin{aligned}
\begin{cases}
\dfrac{1}{\sqrt[4]{\pi}}\sqrt[4]{\dfrac{\left(1+z\right)\left(1-z\right)^3}{z^2}}{\mathcal N}^{3/4}z^{\mathcal N/2}<r_m<1\, ,\\
0<r_m^2-{\mathcal N}^2r_m^2 r_f^2 < 1\, . 
\end{cases} 
\end{aligned}
\end{eqnarray} 
On the contrary, the ALP mass should be smaller than the zero-temperature $Z_{\mathcal N}$ axion mass in the single level crossing.
In fact, we find that the single level crossing can transform into the double level crossings, depending on the masses $m_A$ versus $m_{a,0}$. 

{\it $Z_{\mathcal N}$ axion DM.}
Here we estimate the $Z_{\mathcal N}$ axion relic density in the single and double level crossings through the trapped+kinetic misalignment mechanism \cite{DiLuzio:2021gos, Co:2019jts}.
We consider the pre-inflationary scenario in which the PQ symmetry is spontaneously broken during inflation. 
In order to ensure the adiabatic transition of axion energy density at $T_\times$, we consider a general adiabatic condition that the axions start first to oscillate much earlier than the first level crossing, $\rm i.e.$, 
\begin{eqnarray}
T_{1,a}\gg T_\times\, , \quad T_{1,A}\gg T_\times\, ,
\end{eqnarray}
where $T_1$ is the corresponding oscillation temperature given by $m_a(T)=3H(T)$ and $m_A=3H(T)$ with the Hubble parameter $H(T)$, the subscripts ``$a$" and ``$A$" represent the $Z_{\mathcal N}$ axion and ALP, respectively.

We first discuss the single level crossing case.
At high temperatures, the ALP field is frozen at an arbitrary initial misalignment angle $\theta_{1,A}$, and starts to oscillate at $T_{1,A}$.
The initial energy density in the ALP field is 
\begin{eqnarray}
\rho_{A,1}=\frac{1}{2}m_A^2 f_A^2 \theta_{1,A}^2\, .
\end{eqnarray}
At $T_\times<T<T_{1,A}$, the ALP energy density is adiabatic invariant with the comoving number $N_A \equiv \rho_A a^3 /m_A$, where $a$ is the scale factor.
Then we have the ALP energy density at the level crossing $T_\times$ as
\begin{eqnarray}
\rho_{A,\times}=\frac{1}{2}m_A^2 f_A^2 \theta_{1,A}^2 \left(\frac{a_{1,A}}{a_\times}\right)^3 \, ,
\end{eqnarray}
where $a_{1,A}$ and $a_\times$ are the scale factors at $T_{1,A}$ and $T_\times$, respectively.
At $T=T_\times$, the heavy mass eigenvalue will comprise the $Z_{\mathcal N}$ axion and this energy density $\rho_{A,\times}$ is transferred to the $Z_{\mathcal N}$ axion.
At $T_{\rm QCD}<T<T_\times$, the $Z_{\mathcal N}$ axion is trapped around $\theta_{\rm tr}\equiv\theta_a(T_{\rm QCD})\sim \pi$ until $T_{\rm QCD}$ with the initial axion velocity $\dot{\theta}_{\rm tr}\equiv\dot{\theta}_a(T_{\rm QCD})$.
In this period, the $Z_{\mathcal N}$ axion energy density is adiabatic invariant with $N_a \equiv \rho_a a^3 /m_a$.
Then we have the $Z_{\mathcal N}$ axion energy density at $T_{\rm QCD}$ as
\begin{eqnarray}
\rho_{a, \rm QCD}=\frac{1}{2} m_{a,\pi} m_A f_A^2 \theta_{1,A}^2 \left(\frac{a_{1,A}}{a_{\rm QCD}}\right)^3 \, ,
\end{eqnarray}
where $a_{\rm QCD}$ is the scale factor at $T_{\rm QCD}$.
The mean velocity is given by 
\begin{eqnarray}
\sqrt{\langle\dot{\theta}_{\rm tr}^2\rangle}=\frac{1}{\sqrt {2}} r_f \sqrt{m_{a,\pi} m_A} \theta_{1,A} \left(\frac{a_{1,A}}{a_{\rm QCD}}\right)^{3/2} \, .
\label{mean_velocity_1}
\end{eqnarray}
At $T=T_{\rm QCD}$, the $Z_{\mathcal N}$ axion mass is suddenly exponentially suppressed and the true minimum $\theta_a=0$ develops.
The $Z_{\mathcal N}$ axion energy density at this point is non-adiabatic, which just after $T_{\rm QCD}$ is given by
\begin{eqnarray}
\rho_{a, \rm tr}=\frac{1}{2} f_a^2 \dot{\theta}_{\rm tr}^2 + 2\frac{m_a^2 f_a^2}{{\mathcal N}^2}\, .
\end{eqnarray}
The subsequent period is determined by the value of the velocity $\dot{\theta}_{\rm tr}\sim 2m_a/{\mathcal N}$.
Here we are more concerned about a case that $\dot{\theta}_{\rm tr}\gg 2m_a/{\mathcal N}$, $\rm i.e.$, the kinetic misalignment mechanism will work. 
Then at $T_2<T<T_{\rm QCD}$, the adiabatic approximation is not valid and the $Z_{\mathcal N}$ axion energy density is conserved with the comoving PQ charge $q_{\rm kin}=\dot{\theta}_a a^3$.
The temperature $T_2$ is defined at which the kinetic energy is equal to the barrier height, suggesting $\dot{\theta}_a(T_2)=2m_a/{\mathcal N}$.
Using the conserved $q_{\rm kin}$, we can obtain the scale factor at $T_2$ as $a_2=({\mathcal N}\dot{\theta}_{\rm tr}/(2m_a))^{1/3}a_{\rm QCD}$.
At $T<T_2$, the $Z_{\mathcal N}$ axion will start second to oscillate until nowadays, and the adiabatic approximation is valid again.
Using $\rho_{a,2} a_2^3/m_{a,2}=\rho_{a,0} a_0^3/m_{a,0}$, then we can obtain the $Z_{\mathcal N}$ axion energy density at present
\begin{eqnarray}
\rho_{a,0} = \frac{m_{a,0} f_a^2 \dot{\theta}_{\rm tr}}{\mathcal N} \left(\frac{a_{\rm QCD}}{a_0}\right)^3\, ,
\label{rho_0}
\end{eqnarray}
where $a_0$ is the scale factor today.
Substituting Eq.~(\ref{mean_velocity_1}) in above equation, $\rho_{a,0}$ reads
\begin{eqnarray}
\rho_{a,0} = C \frac{r_f m_{a,0} \sqrt{m_{a,\pi} m_A} \theta_{1,A} f_a^2}{\sqrt{2}{\mathcal N}} \left(\frac{\sqrt{a_{1,A} a_{\rm QCD}}}{a_0}\right)^3\, , ~~~
\end{eqnarray}
where $C\simeq2$ is a constant.
Compared with the no level crossing case \cite{DiLuzio:2021gos}, the ratio of the $Z_{\mathcal N}$ axion relic density is given by
\begin{eqnarray}
\begin{aligned}
\frac{\rho_{a,0}^{\rm (slc)}}{\rho_{a,0}^{\rm (no)}}&=r_f \frac{\sqrt {m_A}}{\sqrt {m_{a,1}}} \frac{\theta_{1,A}}{|\theta'_{1,a}-\pi|} \left(\frac{\sqrt{a_{1,A}}}{\sqrt{a_{1,a}}}\right)^3\\
&\simeq r_f \left(\frac{m_{a,1}}{m_A}\right)^{1/4} \frac{\theta_{1,A}}{|\theta'_{1,a}-\pi|}\, ,
\end{aligned}
\end{eqnarray}
where $m_{a,1}$, $\theta'_{1,a}$, and $a_{1,a}$ correspond to the $Z_{\mathcal N}$ axion  oscillating at $T_{1,a}$ in the no level crossing case.
Since $r_f\ll 1$, $m_{a,1}\ll m_A$, and $|\theta'_{1,a}-\pi| \sim \pi$, we have the ratio $\rho_{a,0}^{\rm (slc)}/\rho_{a,0}^{\rm (no)}\ll 1$. 
Therefore, the $Z_{\mathcal N}$ axion relic density is significantly suppressed in the single level crossing. 
 
Then we discuss the double level crossings case. 
At high temperatures, the $Z_{\mathcal N}$ axion starts to oscillate at $T_{1,a}$ with the initial misalignment angle $\theta_{1,a}$.
The initial energy density is 
\begin{eqnarray}
\rho_{a,1}=\frac{1}{2}m_{a,1}^2 f_a^2 \theta_{1,a}^2\, .
\end{eqnarray}
At $T_\times<T<T_{1,A}$, the $Z_{\mathcal N}$ axion energy density is adiabatic invariant, which at $T_\times$ is given by
\begin{eqnarray}
\rho_{a,\times}=\frac{1}{2} m_A m_{a,1} f_a^2 \theta_{1,a}^2 \left(\frac{a_{1,a}}{a_\times}\right)^3 \, .
\end{eqnarray}
At the first level crossing $T=T_\times$, the light mass eigenvalue will comprise the ALP and this energy density is transferred to the ALP.
At $T_{\rm QCD}<T<T_\times$, the adiabatic approximation is valid, we have the ALP energy density at $T_{\rm QCD}$ as
\begin{eqnarray}
\rho_{A, \rm QCD}=\frac{1}{2} m_A m_{a,1} f_a^2 \theta_{1,a}^2 \left(\frac{a_{1,a}}{a_{\rm QCD}}\right)^3 \, ,
\end{eqnarray}
and the mean velocity as 
\begin{eqnarray}
\sqrt{\langle\dot{\theta}_{\rm tr}^2\rangle}=\frac{1}{\sqrt {2}} \frac{1}{r_f} \sqrt{m_A m_{a,1}} \theta_{1,a} \left(\frac{a_{1,a}}{a_{\rm QCD}}\right)^{3/2} \, .
\label{mean_velocity_2}
\end{eqnarray}
At $T=T_{\rm QCD}$, the second level crossing occurs and the light mass eigenvalue will comprise the $Z_{\mathcal N}$ axion again.
Note that the $Z_{\mathcal N}$ axion energy density at the second level crossing is non-adiabatic, and the subsequent period ($T<T_{\rm QCD}$) is similar to the single level crossing case.
By substituting Eq.~(\ref{mean_velocity_2}) in Eq.~(\ref{rho_0}), we have the present $Z_{\mathcal N}$ axion energy density 
\begin{eqnarray}
\rho_{a,0} = C' \frac{m_{a,0} \sqrt{m_A m_{a,1}} \theta_{1,a} f_a^2}{\sqrt{2}{\mathcal N}r_f} \left(\frac{\sqrt{a_{1,a} a_{\rm QCD}}}{a_0}\right)^3\, . ~
\end{eqnarray} 
Then the ratio of the relic density in this case is given by
\begin{eqnarray}
\begin{aligned}
\frac{\rho_{a,0}^{\rm (dlc)}}{\rho_{a,0}^{\rm (no)}}&=\frac{1}{r_f} \frac{\sqrt {m_A}}{\sqrt {m_{a,\pi}}} \frac{\theta_{1,a}}{|\theta'_{1,a}-\pi|} \\
&= \frac{1}{r_f} \sqrt{r_m} \frac{\theta_{1,a}}{|\theta'_{1,a}-\pi|}\, .
\end{aligned}
\end{eqnarray} 
Since $r_m\ll 1$, the $Z_{\mathcal N}$ axion relic density can be enhanced or suppressed in the double level crossings. 
   
{\it Conclusion.}
In summary, we have investigated the level crossing between the $Z_{\mathcal N}$ axion and ALP, and found that the single and double level crossings can naturally take place in their mass mixing.
The first level crossing occurs much earlier than the QCD phase transition, while the second level crossing occurs exactly during the QCD phase transition if it exists.
The condition for the double level crossings to occur is that, 1) the first level crossing should take place before; 2) the ALP mass should be larger than the zero-temperature $Z_{\mathcal N}$ axion mass. 
The cosmological evolution of axions will become significantly different due to the presence of the second level crossing.
In the single level crossing, the light mass eigenvalue comprises the $Z_{\mathcal N}$ axion at high temperatures. 
When the temperature is lower than the level crossing temperature $T_\times$, the $Z_{\mathcal N}$ axion is comprised by the heavy mass eigenvalue.
In the double level crossings, the axion evolution is similar to the single level crossing case at high temperatures $T>T_{\rm QCD}$.
However, when the temperature is below $T_{\rm QCD}$, the light mass eigenvalue will comprise the $Z_{\mathcal N}$ axion again.

Comparing the single level crossing with the double level crossings, the main distinction depends on the relation between the ALP mass and the zero-temperature $Z_{\mathcal N}$ axion mass $m_A \sim m_{a,0}$.
%\begin{eqnarray}
%m_A \sim m_{a,0}\, .
%\end{eqnarray}
If $m_A < m_{a,0}$, there is the single level crossing.
On the contrary, once $m_A > m_{a,0}$, the single level crossing will naturally transform into the double level crossings.
We find that this single to double level crossings transformation only exists in the mass mixing between the $Z_{\mathcal N}$ axion and ALP, which is the main difference from the canonical level crossing between the QCD axion and ALP.

Finally, we estimate the current $Z_{\mathcal N}$ axion relic density in the single and double level crossings through the trapped+kinetic misalignment mechanism.
Compared with the no level crossing case, we find that the $Z_{\mathcal N}$ axion relic density can be suppressed in the single level crossing case, and enhanced or suppressed in the double level crossings case.

{\bf Acknowledgments.}
The authors would like to thank David Cyncynates for helpful discussions and valuable comments.
W.C. is supported by the National Natural Science Foundation of China (NSFC) (Grants No.~11775025 and No.~12175027).
Y.F.Z. is supported by the National Key R\&D Program of China (Grant No.~2017YFA0402204), the CAS Project for Young Scientists in Basic Research YSBR-006, and the NSFC (Grants No.~11821505, No.~11825506, and No.~12047503).

\bibliography{references}

%merlin.mbs apsrev4-1.bst 2010-07-25 4.21a (PWD, AO, DPC) hacked
%Control: key (0)
%Control: author (0) dotless jnrlst
%Control: editor formatted (1) identically to author
%Control: production of article title (0) allowed
%Control: page (1) range
%Control: year (0) verbatim
%Control: production of eprint (0) enabled
\begin{thebibliography}{31}%
\makeatletter
\providecommand \@ifxundefined [1]{%
 \@ifx{#1\undefined}
}%
\providecommand \@ifnum [1]{%
 \ifnum #1\expandafter \@firstoftwo
 \else \expandafter \@secondoftwo
 \fi
}%
\providecommand \@ifx [1]{%
 \ifx #1\expandafter \@firstoftwo
 \else \expandafter \@secondoftwo
 \fi
}%
\providecommand \natexlab [1]{#1}%
\providecommand \enquote  [1]{``#1''}%
\providecommand \bibnamefont  [1]{#1}%
\providecommand \bibfnamefont [1]{#1}%
\providecommand \citenamefont [1]{#1}%
\providecommand \href@noop [0]{\@secondoftwo}%
\providecommand \href [0]{\begingroup \@sanitize@url \@href}%
\providecommand \@href[1]{\@@startlink{#1}\@@href}%
\providecommand \@@href[1]{\endgroup#1\@@endlink}%
\providecommand \@sanitize@url [0]{\catcode `\\12\catcode `\$12\catcode
  `\&12\catcode `\#12\catcode `\^12\catcode `\_12\catcode `\%12\relax}%
\providecommand \@@startlink[1]{}%
\providecommand \@@endlink[0]{}%
\providecommand \url  [0]{\begingroup\@sanitize@url \@url }%
\providecommand \@url [1]{\endgroup\@href {#1}{\urlprefix }}%
\providecommand \urlprefix  [0]{URL }%
\providecommand \Eprint [0]{\href }%
\providecommand \doibase [0]{http://dx.doi.org/}%
\providecommand \selectlanguage [0]{\@gobble}%
\providecommand \bibinfo  [0]{\@secondoftwo}%
\providecommand \bibfield  [0]{\@secondoftwo}%
\providecommand \translation [1]{[#1]}%
\providecommand \BibitemOpen [0]{}%
\providecommand \bibitemStop [0]{}%
\providecommand \bibitemNoStop [0]{.\EOS\space}%
\providecommand \EOS [0]{\spacefactor3000\relax}%
\providecommand \BibitemShut  [1]{\csname bibitem#1\endcsname}%
\let\auto@bib@innerbib\@empty
%</preamble>
\bibitem [{\citenamefont {Peccei}\ and\ \citenamefont
  {Quinn}(1977{\natexlab{a}})}]{Peccei:1977ur}%
  \BibitemOpen
  \bibfield  {author} {\bibinfo {author} {\bibfnamefont {R.D.}\ \bibnamefont
  {Peccei}}\ and\ \bibinfo {author} {\bibfnamefont {Helen~R.}\ \bibnamefont
  {Quinn}},\ }\bibfield  {title} {\enquote {\bibinfo {title} {{Constraints
  Imposed by CP Conservation in the Presence of Instantons}},}\ }\href
  {\doibase 10.1103/PhysRevD.16.1791} {\bibfield  {journal} {\bibinfo
  {journal} {Phys. Rev. D}\ }\textbf {\bibinfo {volume} {16}},\ \bibinfo
  {pages} {1791--1797} (\bibinfo {year} {1977}{\natexlab{a}})}\BibitemShut
  {NoStop}%
\bibitem [{\citenamefont {Peccei}\ and\ \citenamefont
  {Quinn}(1977{\natexlab{b}})}]{Peccei:1977hh}%
  \BibitemOpen
  \bibfield  {author} {\bibinfo {author} {\bibfnamefont {R.D.}\ \bibnamefont
  {Peccei}}\ and\ \bibinfo {author} {\bibfnamefont {Helen~R.}\ \bibnamefont
  {Quinn}},\ }\bibfield  {title} {\enquote {\bibinfo {title} {{CP Conservation
  in the Presence of Instantons}},}\ }\href {\doibase
  10.1103/PhysRevLett.38.1440} {\bibfield  {journal} {\bibinfo  {journal}
  {Phys. Rev. Lett.}\ }\textbf {\bibinfo {volume} {38}},\ \bibinfo {pages}
  {1440--1443} (\bibinfo {year} {1977}{\natexlab{b}})}\BibitemShut {NoStop}%
\bibitem [{\citenamefont {Weinberg}(1978)}]{Weinberg:1977ma}%
  \BibitemOpen
  \bibfield  {author} {\bibinfo {author} {\bibfnamefont {Steven}\ \bibnamefont
  {Weinberg}},\ }\bibfield  {title} {\enquote {\bibinfo {title} {{A New Light
  Boson?}}}\ }\href {\doibase 10.1103/PhysRevLett.40.223} {\bibfield  {journal}
  {\bibinfo  {journal} {Phys. Rev. Lett.}\ }\textbf {\bibinfo {volume} {40}},\
  \bibinfo {pages} {223--226} (\bibinfo {year} {1978})}\BibitemShut {NoStop}%
\bibitem [{\citenamefont {Wilczek}(1978)}]{Wilczek:1977pj}%
  \BibitemOpen
  \bibfield  {author} {\bibinfo {author} {\bibfnamefont {Frank}\ \bibnamefont
  {Wilczek}},\ }\bibfield  {title} {\enquote {\bibinfo {title} {{Problem of
  Strong $P$ and $T$ Invariance in the Presence of Instantons}},}\ }\href
  {\doibase 10.1103/PhysRevLett.40.279} {\bibfield  {journal} {\bibinfo
  {journal} {Phys. Rev. Lett.}\ }\textbf {\bibinfo {volume} {40}},\ \bibinfo
  {pages} {279--282} (\bibinfo {year} {1978})}\BibitemShut {NoStop}%
\bibitem [{\citenamefont {Kim}(1979)}]{Kim:1979if}%
  \BibitemOpen
  \bibfield  {author} {\bibinfo {author} {\bibfnamefont {Jihn~E.}\ \bibnamefont
  {Kim}},\ }\bibfield  {title} {\enquote {\bibinfo {title} {{Weak Interaction
  Singlet and Strong CP Invariance}},}\ }\href {\doibase
  10.1103/PhysRevLett.43.103} {\bibfield  {journal} {\bibinfo  {journal} {Phys.
  Rev. Lett.}\ }\textbf {\bibinfo {volume} {43}},\ \bibinfo {pages} {103}
  (\bibinfo {year} {1979})}\BibitemShut {NoStop}%
\bibitem [{\citenamefont {Shifman}\ \emph {et~al.}(1980)\citenamefont
  {Shifman}, \citenamefont {Vainshtein},\ and\ \citenamefont
  {Zakharov}}]{Shifman:1979if}%
  \BibitemOpen
  \bibfield  {author} {\bibinfo {author} {\bibfnamefont {Mikhail~A.}\
  \bibnamefont {Shifman}}, \bibinfo {author} {\bibfnamefont {A.~I.}\
  \bibnamefont {Vainshtein}}, \ and\ \bibinfo {author} {\bibfnamefont
  {Valentin~I.}\ \bibnamefont {Zakharov}},\ }\bibfield  {title} {\enquote
  {\bibinfo {title} {{Can Confinement Ensure Natural CP Invariance of Strong
  Interactions?}}}\ }\href {\doibase 10.1016/0550-3213(80)90209-6} {\bibfield
  {journal} {\bibinfo  {journal} {Nucl. Phys. B}\ }\textbf {\bibinfo {volume}
  {166}},\ \bibinfo {pages} {493--506} (\bibinfo {year} {1980})}\BibitemShut
  {NoStop}%
\bibitem [{\citenamefont {Dine}\ \emph {et~al.}(1981)\citenamefont {Dine},
  \citenamefont {Fischler},\ and\ \citenamefont {Srednicki}}]{Dine:1981rt}%
  \BibitemOpen
  \bibfield  {author} {\bibinfo {author} {\bibfnamefont {Michael}\ \bibnamefont
  {Dine}}, \bibinfo {author} {\bibfnamefont {Willy}\ \bibnamefont {Fischler}},
  \ and\ \bibinfo {author} {\bibfnamefont {Mark}\ \bibnamefont {Srednicki}},\
  }\bibfield  {title} {\enquote {\bibinfo {title} {{A Simple Solution to the
  Strong CP Problem with a Harmless Axion}},}\ }\href {\doibase
  10.1016/0370-2693(81)90590-6} {\bibfield  {journal} {\bibinfo  {journal}
  {Phys. Lett. B}\ }\textbf {\bibinfo {volume} {104}},\ \bibinfo {pages}
  {199--202} (\bibinfo {year} {1981})}\BibitemShut {NoStop}%
\bibitem [{\citenamefont {Zhitnitsky}(1980)}]{Zhitnitsky:1980tq}%
  \BibitemOpen
  \bibfield  {author} {\bibinfo {author} {\bibfnamefont {A.~R.}\ \bibnamefont
  {Zhitnitsky}},\ }\bibfield  {title} {\enquote {\bibinfo {title} {{On Possible
  Suppression of the Axion Hadron Interactions. (In Russian)}},}\ }\href@noop
  {} {\bibfield  {journal} {\bibinfo  {journal} {Sov. J. Nucl. Phys.}\ }\textbf
  {\bibinfo {volume} {31}},\ \bibinfo {pages} {260} (\bibinfo {year}
  {1980})}\BibitemShut {NoStop}%
\bibitem [{\citenamefont {'t~Hooft}(1976{\natexlab{a}})}]{tHooft:1976rip}%
  \BibitemOpen
  \bibfield  {author} {\bibinfo {author} {\bibfnamefont {Gerard}\ \bibnamefont
  {'t~Hooft}},\ }\bibfield  {title} {\enquote {\bibinfo {title} {{Symmetry
  Breaking Through Bell-Jackiw Anomalies}},}\ }\href {\doibase
  10.1103/PhysRevLett.37.8} {\bibfield  {journal} {\bibinfo  {journal} {Phys.
  Rev. Lett.}\ }\textbf {\bibinfo {volume} {37}},\ \bibinfo {pages} {8--11}
  (\bibinfo {year} {1976}{\natexlab{a}})}\BibitemShut {NoStop}%
\bibitem [{\citenamefont {'t~Hooft}(1976{\natexlab{b}})}]{tHooft:1976snw}%
  \BibitemOpen
  \bibfield  {author} {\bibinfo {author} {\bibfnamefont {Gerard}\ \bibnamefont
  {'t~Hooft}},\ }\bibfield  {title} {\enquote {\bibinfo {title} {{Computation
  of the Quantum Effects Due to a Four-Dimensional Pseudoparticle}},}\ }\href
  {\doibase 10.1103/PhysRevD.14.3432} {\bibfield  {journal} {\bibinfo
  {journal} {Phys. Rev. D}\ }\textbf {\bibinfo {volume} {14}},\ \bibinfo
  {pages} {3432--3450} (\bibinfo {year} {1976}{\natexlab{b}})},\ \bibinfo
  {note} {[Erratum: Phys.Rev.D 18, 2199 (1978)]}\BibitemShut {NoStop}%
\bibitem [{\citenamefont {Preskill}\ \emph {et~al.}(1983)\citenamefont
  {Preskill}, \citenamefont {Wise},\ and\ \citenamefont
  {Wilczek}}]{Preskill:1982cy}%
  \BibitemOpen
  \bibfield  {author} {\bibinfo {author} {\bibfnamefont {John}\ \bibnamefont
  {Preskill}}, \bibinfo {author} {\bibfnamefont {Mark~B.}\ \bibnamefont
  {Wise}}, \ and\ \bibinfo {author} {\bibfnamefont {Frank}\ \bibnamefont
  {Wilczek}},\ }\bibfield  {title} {\enquote {\bibinfo {title} {{Cosmology of
  the Invisible Axion}},}\ }\href {\doibase 10.1016/0370-2693(83)90637-8}
  {\bibfield  {journal} {\bibinfo  {journal} {Phys. Lett. B}\ }\textbf
  {\bibinfo {volume} {120}},\ \bibinfo {pages} {127--132} (\bibinfo {year}
  {1983})}\BibitemShut {NoStop}%
\bibitem [{\citenamefont {Abbott}\ and\ \citenamefont
  {Sikivie}(1983)}]{Abbott:1982af}%
  \BibitemOpen
  \bibfield  {author} {\bibinfo {author} {\bibfnamefont {L.F.}\ \bibnamefont
  {Abbott}}\ and\ \bibinfo {author} {\bibfnamefont {P.}~\bibnamefont
  {Sikivie}},\ }\bibfield  {title} {\enquote {\bibinfo {title} {{A Cosmological
  Bound on the Invisible Axion}},}\ }\href {\doibase
  10.1016/0370-2693(83)90638-X} {\bibfield  {journal} {\bibinfo  {journal}
  {Phys. Lett. B}\ }\textbf {\bibinfo {volume} {120}},\ \bibinfo {pages}
  {133--136} (\bibinfo {year} {1983})}\BibitemShut {NoStop}%
\bibitem [{\citenamefont {Dine}\ and\ \citenamefont
  {Fischler}(1983)}]{Dine:1982ah}%
  \BibitemOpen
  \bibfield  {author} {\bibinfo {author} {\bibfnamefont {Michael}\ \bibnamefont
  {Dine}}\ and\ \bibinfo {author} {\bibfnamefont {Willy}\ \bibnamefont
  {Fischler}},\ }\bibfield  {title} {\enquote {\bibinfo {title} {{The Not So
  Harmless Axion}},}\ }\href {\doibase 10.1016/0370-2693(83)90639-1} {\bibfield
   {journal} {\bibinfo  {journal} {Phys. Lett. B}\ }\textbf {\bibinfo {volume}
  {120}},\ \bibinfo {pages} {137--141} (\bibinfo {year} {1983})}\BibitemShut
  {NoStop}%
\bibitem [{\citenamefont {Hook}(2018)}]{Hook:2018jle}%
  \BibitemOpen
  \bibfield  {author} {\bibinfo {author} {\bibfnamefont {Anson}\ \bibnamefont
  {Hook}},\ }\bibfield  {title} {\enquote {\bibinfo {title} {{Solving the
  Hierarchy Problem Discretely}},}\ }\href {\doibase
  10.1103/PhysRevLett.120.261802} {\bibfield  {journal} {\bibinfo  {journal}
  {Phys. Rev. Lett.}\ }\textbf {\bibinfo {volume} {120}},\ \bibinfo {pages}
  {261802} (\bibinfo {year} {2018})},\ \Eprint
  {http://arxiv.org/abs/1802.10093} {arXiv:1802.10093 [hep-ph]} \BibitemShut
  {NoStop}%
\bibitem [{\citenamefont {Di~Luzio}\ \emph
  {et~al.}(2021{\natexlab{a}})\citenamefont {Di~Luzio}, \citenamefont {Gavela},
  \citenamefont {Quilez},\ and\ \citenamefont {Ringwald}}]{DiLuzio:2021pxd}%
  \BibitemOpen
  \bibfield  {author} {\bibinfo {author} {\bibfnamefont {Luca}\ \bibnamefont
  {Di~Luzio}}, \bibinfo {author} {\bibfnamefont {Belen}\ \bibnamefont
  {Gavela}}, \bibinfo {author} {\bibfnamefont {Pablo}\ \bibnamefont {Quilez}},
  \ and\ \bibinfo {author} {\bibfnamefont {Andreas}\ \bibnamefont {Ringwald}},\
  }\bibfield  {title} {\enquote {\bibinfo {title} {{An even lighter QCD
  axion}},}\ }\href {\doibase 10.1007/JHEP05(2021)184} {\bibfield  {journal}
  {\bibinfo  {journal} {JHEP}\ }\textbf {\bibinfo {volume} {05}},\ \bibinfo
  {pages} {184} (\bibinfo {year} {2021}{\natexlab{a}})},\ \Eprint
  {http://arxiv.org/abs/2102.00012} {arXiv:2102.00012 [hep-ph]} \BibitemShut
  {NoStop}%
\bibitem [{\citenamefont {Di~Luzio}\ \emph
  {et~al.}(2021{\natexlab{b}})\citenamefont {Di~Luzio}, \citenamefont {Gavela},
  \citenamefont {Quilez},\ and\ \citenamefont {Ringwald}}]{DiLuzio:2021gos}%
  \BibitemOpen
  \bibfield  {author} {\bibinfo {author} {\bibfnamefont {Luca}\ \bibnamefont
  {Di~Luzio}}, \bibinfo {author} {\bibfnamefont {Belen}\ \bibnamefont
  {Gavela}}, \bibinfo {author} {\bibfnamefont {Pablo}\ \bibnamefont {Quilez}},
  \ and\ \bibinfo {author} {\bibfnamefont {Andreas}\ \bibnamefont {Ringwald}},\
  }\bibfield  {title} {\enquote {\bibinfo {title} {{Dark matter from an even
  lighter QCD axion: trapped misalignment}},}\ }\href {\doibase
  10.1088/1475-7516/2021/10/001} {\bibfield  {journal} {\bibinfo  {journal}
  {JCAP}\ }\textbf {\bibinfo {volume} {10}},\ \bibinfo {pages} {001} (\bibinfo
  {year} {2021}{\natexlab{b}})},\ \Eprint {http://arxiv.org/abs/2102.01082}
  {arXiv:2102.01082 [hep-ph]} \BibitemShut {NoStop}%
\bibitem [{\citenamefont {Co}\ \emph {et~al.}(2020)\citenamefont {Co},
  \citenamefont {Hall},\ and\ \citenamefont {Harigaya}}]{Co:2019jts}%
  \BibitemOpen
  \bibfield  {author} {\bibinfo {author} {\bibfnamefont {Raymond~T.}\
  \bibnamefont {Co}}, \bibinfo {author} {\bibfnamefont {Lawrence~J.}\
  \bibnamefont {Hall}}, \ and\ \bibinfo {author} {\bibfnamefont {Keisuke}\
  \bibnamefont {Harigaya}},\ }\bibfield  {title} {\enquote {\bibinfo {title}
  {{Axion Kinetic Misalignment Mechanism}},}\ }\href {\doibase
  10.1103/PhysRevLett.124.251802} {\bibfield  {journal} {\bibinfo  {journal}
  {Phys. Rev. Lett.}\ }\textbf {\bibinfo {volume} {124}},\ \bibinfo {pages}
  {251802} (\bibinfo {year} {2020})},\ \Eprint
  {http://arxiv.org/abs/1910.14152} {arXiv:1910.14152 [hep-ph]} \BibitemShut
  {NoStop}%
\bibitem [{\citenamefont {Green}\ and\ \citenamefont
  {Schwarz}(1984)}]{Green:1984sg}%
  \BibitemOpen
  \bibfield  {author} {\bibinfo {author} {\bibfnamefont {Michael~B.}\
  \bibnamefont {Green}}\ and\ \bibinfo {author} {\bibfnamefont {John~H.}\
  \bibnamefont {Schwarz}},\ }\bibfield  {title} {\enquote {\bibinfo {title}
  {{Anomaly Cancellation in Supersymmetric D=10 Gauge Theory and Superstring
  Theory}},}\ }\href {\doibase 10.1016/0370-2693(84)91565-X} {\bibfield
  {journal} {\bibinfo  {journal} {Phys. Lett. B}\ }\textbf {\bibinfo {volume}
  {149}},\ \bibinfo {pages} {117--122} (\bibinfo {year} {1984})}\BibitemShut
  {NoStop}%
\bibitem [{\citenamefont {Witten}(1984)}]{Witten:1984dg}%
  \BibitemOpen
  \bibfield  {author} {\bibinfo {author} {\bibfnamefont {Edward}\ \bibnamefont
  {Witten}},\ }\bibfield  {title} {\enquote {\bibinfo {title} {{Some Properties
  of O(32) Superstrings}},}\ }\href {\doibase 10.1016/0370-2693(84)90422-2}
  {\bibfield  {journal} {\bibinfo  {journal} {Phys. Lett. B}\ }\textbf
  {\bibinfo {volume} {149}},\ \bibinfo {pages} {351--356} (\bibinfo {year}
  {1984})}\BibitemShut {NoStop}%
\bibitem [{\citenamefont {Cadamuro}\ and\ \citenamefont
  {Redondo}(2012)}]{Cadamuro:2011fd}%
  \BibitemOpen
  \bibfield  {author} {\bibinfo {author} {\bibfnamefont {Davide}\ \bibnamefont
  {Cadamuro}}\ and\ \bibinfo {author} {\bibfnamefont {Javier}\ \bibnamefont
  {Redondo}},\ }\bibfield  {title} {\enquote {\bibinfo {title} {{Cosmological
  bounds on pseudo Nambu-Goldstone bosons}},}\ }\href {\doibase
  10.1088/1475-7516/2012/02/032} {\bibfield  {journal} {\bibinfo  {journal}
  {JCAP}\ }\textbf {\bibinfo {volume} {02}},\ \bibinfo {pages} {032} (\bibinfo
  {year} {2012})},\ \Eprint {http://arxiv.org/abs/1110.2895} {arXiv:1110.2895
  [hep-ph]} \BibitemShut {NoStop}%
\bibitem [{\citenamefont {Arias}\ \emph {et~al.}(2012)\citenamefont {Arias},
  \citenamefont {Cadamuro}, \citenamefont {Goodsell}, \citenamefont {Jaeckel},
  \citenamefont {Redondo},\ and\ \citenamefont {Ringwald}}]{Arias:2012az}%
  \BibitemOpen
  \bibfield  {author} {\bibinfo {author} {\bibfnamefont {Paola}\ \bibnamefont
  {Arias}}, \bibinfo {author} {\bibfnamefont {Davide}\ \bibnamefont
  {Cadamuro}}, \bibinfo {author} {\bibfnamefont {Mark}\ \bibnamefont
  {Goodsell}}, \bibinfo {author} {\bibfnamefont {Joerg}\ \bibnamefont
  {Jaeckel}}, \bibinfo {author} {\bibfnamefont {Javier}\ \bibnamefont
  {Redondo}}, \ and\ \bibinfo {author} {\bibfnamefont {Andreas}\ \bibnamefont
  {Ringwald}},\ }\bibfield  {title} {\enquote {\bibinfo {title} {{WISPy Cold
  Dark Matter}},}\ }\href {\doibase 10.1088/1475-7516/2012/06/013} {\bibfield
  {journal} {\bibinfo  {journal} {JCAP}\ }\textbf {\bibinfo {volume} {06}},\
  \bibinfo {pages} {013} (\bibinfo {year} {2012})},\ \Eprint
  {http://arxiv.org/abs/1201.5902} {arXiv:1201.5902 [hep-ph]} \BibitemShut
  {NoStop}%
\bibitem [{\citenamefont {Chao}\ \emph {et~al.}(2022)\citenamefont {Chao},
  \citenamefont {Jin}, \citenamefont {Li},\ and\ \citenamefont
  {Peng}}]{Chao:2022blc}%
  \BibitemOpen
  \bibfield  {author} {\bibinfo {author} {\bibfnamefont {Wei}\ \bibnamefont
  {Chao}}, \bibinfo {author} {\bibfnamefont {Mingjie}\ \bibnamefont {Jin}},
  \bibinfo {author} {\bibfnamefont {Hai-Jun}\ \bibnamefont {Li}}, \ and\
  \bibinfo {author} {\bibfnamefont {Ying-Quan}\ \bibnamefont {Peng}},\
  }\bibfield  {title} {\enquote {\bibinfo {title} {{Axion-like Dark Matter from
  the Type-II Seesaw Mechanism}},}\ }\href@noop {} {\  (\bibinfo {year}
  {2022})},\ \Eprint {http://arxiv.org/abs/2210.13233} {arXiv:2210.13233
  [hep-ph]} \BibitemShut {NoStop}%
\bibitem [{\citenamefont {Hill}\ and\ \citenamefont
  {Ross}(1988)}]{Hill:1988bu}%
  \BibitemOpen
  \bibfield  {author} {\bibinfo {author} {\bibfnamefont {Christopher~T.}\
  \bibnamefont {Hill}}\ and\ \bibinfo {author} {\bibfnamefont {Graham~G.}\
  \bibnamefont {Ross}},\ }\bibfield  {title} {\enquote {\bibinfo {title}
  {{Models and New Phenomenological Implications of a Class of Pseudogoldstone
  Bosons}},}\ }\href {\doibase 10.1016/0550-3213(88)90062-4} {\bibfield
  {journal} {\bibinfo  {journal} {Nucl. Phys. B}\ }\textbf {\bibinfo {volume}
  {311}},\ \bibinfo {pages} {253--297} (\bibinfo {year} {1988})}\BibitemShut
  {NoStop}%
\bibitem [{\citenamefont {Kitajima}\ and\ \citenamefont
  {Takahashi}(2015)}]{Kitajima:2014xla}%
  \BibitemOpen
  \bibfield  {author} {\bibinfo {author} {\bibfnamefont {Naoya}\ \bibnamefont
  {Kitajima}}\ and\ \bibinfo {author} {\bibfnamefont {Fuminobu}\ \bibnamefont
  {Takahashi}},\ }\bibfield  {title} {\enquote {\bibinfo {title} {{Resonant
  conversions of QCD axions into hidden axions and suppressed isocurvature
  perturbations}},}\ }\href {\doibase 10.1088/1475-7516/2015/01/032} {\bibfield
   {journal} {\bibinfo  {journal} {JCAP}\ }\textbf {\bibinfo {volume} {01}},\
  \bibinfo {pages} {032} (\bibinfo {year} {2015})},\ \Eprint
  {http://arxiv.org/abs/1411.2011} {arXiv:1411.2011 [hep-ph]} \BibitemShut
  {NoStop}%
\bibitem [{\citenamefont {Daido}\ \emph {et~al.}(2016)\citenamefont {Daido},
  \citenamefont {Kitajima},\ and\ \citenamefont {Takahashi}}]{Daido:2015cba}%
  \BibitemOpen
  \bibfield  {author} {\bibinfo {author} {\bibfnamefont {Ryuji}\ \bibnamefont
  {Daido}}, \bibinfo {author} {\bibfnamefont {Naoya}\ \bibnamefont {Kitajima}},
  \ and\ \bibinfo {author} {\bibfnamefont {Fuminobu}\ \bibnamefont
  {Takahashi}},\ }\bibfield  {title} {\enquote {\bibinfo {title} {{Level
  crossing between the QCD axion and an axionlike particle}},}\ }\href
  {\doibase 10.1103/PhysRevD.93.075027} {\bibfield  {journal} {\bibinfo
  {journal} {Phys. Rev. D}\ }\textbf {\bibinfo {volume} {93}},\ \bibinfo
  {pages} {075027} (\bibinfo {year} {2016})},\ \Eprint
  {http://arxiv.org/abs/1510.06675} {arXiv:1510.06675 [hep-ph]} \BibitemShut
  {NoStop}%
\bibitem [{\citenamefont {Daido}\ \emph {et~al.}(2015)\citenamefont {Daido},
  \citenamefont {Kitajima},\ and\ \citenamefont {Takahashi}}]{Daido:2015bva}%
  \BibitemOpen
  \bibfield  {author} {\bibinfo {author} {\bibfnamefont {Ryuji}\ \bibnamefont
  {Daido}}, \bibinfo {author} {\bibfnamefont {Naoya}\ \bibnamefont {Kitajima}},
  \ and\ \bibinfo {author} {\bibfnamefont {Fuminobu}\ \bibnamefont
  {Takahashi}},\ }\bibfield  {title} {\enquote {\bibinfo {title} {{Domain Wall
  Formation from Level Crossing in the Axiverse}},}\ }\href {\doibase
  10.1103/PhysRevD.92.063512} {\bibfield  {journal} {\bibinfo  {journal} {Phys.
  Rev. D}\ }\textbf {\bibinfo {volume} {92}},\ \bibinfo {pages} {063512}
  (\bibinfo {year} {2015})},\ \Eprint {http://arxiv.org/abs/1505.07670}
  {arXiv:1505.07670 [hep-ph]} \BibitemShut {NoStop}%
\bibitem [{\citenamefont {Ho}\ \emph {et~al.}(2018)\citenamefont {Ho},
  \citenamefont {Saikawa},\ and\ \citenamefont {Takahashi}}]{Ho:2018qur}%
  \BibitemOpen
  \bibfield  {author} {\bibinfo {author} {\bibfnamefont {Shu-Yu}\ \bibnamefont
  {Ho}}, \bibinfo {author} {\bibfnamefont {Ken'ichi}\ \bibnamefont {Saikawa}},
  \ and\ \bibinfo {author} {\bibfnamefont {Fuminobu}\ \bibnamefont
  {Takahashi}},\ }\bibfield  {title} {\enquote {\bibinfo {title} {{Enhanced
  photon coupling of ALP dark matter adiabatically converted from the QCD
  axion}},}\ }\href {\doibase 10.1088/1475-7516/2018/10/042} {\bibfield
  {journal} {\bibinfo  {journal} {JCAP}\ }\textbf {\bibinfo {volume} {10}},\
  \bibinfo {pages} {042} (\bibinfo {year} {2018})},\ \Eprint
  {http://arxiv.org/abs/1806.09551} {arXiv:1806.09551 [hep-ph]} \BibitemShut
  {NoStop}%
\bibitem [{\citenamefont {Gavela}\ \emph {et~al.}(2023)\citenamefont {Gavela},
  \citenamefont {Qu\'\i{}lez},\ and\ \citenamefont {Ramos}}]{Gavela:2023tzu}%
  \BibitemOpen
  \bibfield  {author} {\bibinfo {author} {\bibfnamefont {Bel\'en}\ \bibnamefont
  {Gavela}}, \bibinfo {author} {\bibfnamefont {Pablo}\ \bibnamefont
  {Qu\'\i{}lez}}, \ and\ \bibinfo {author} {\bibfnamefont {Maria}\ \bibnamefont
  {Ramos}},\ }\bibfield  {title} {\enquote {\bibinfo {title} {{The QCD axion
  sum rule}},}\ }\href@noop {} {\  (\bibinfo {year} {2023})},\ \Eprint
  {http://arxiv.org/abs/2305.15465} {arXiv:2305.15465 [hep-ph]} \BibitemShut
  {NoStop}%
\bibitem [{\citenamefont {Murai}\ \emph {et~al.}(2023)\citenamefont {Murai},
  \citenamefont {Takahashi},\ and\ \citenamefont {Yin}}]{Murai:2023xjn}%
  \BibitemOpen
  \bibfield  {author} {\bibinfo {author} {\bibfnamefont {Kai}\ \bibnamefont
  {Murai}}, \bibinfo {author} {\bibfnamefont {Fuminobu}\ \bibnamefont
  {Takahashi}}, \ and\ \bibinfo {author} {\bibfnamefont {Wen}\ \bibnamefont
  {Yin}},\ }\bibfield  {title} {\enquote {\bibinfo {title} {{QCD axion: A
  unique player in the axiverse with mixings}},}\ }\href {\doibase
  10.1103/PhysRevD.108.036020} {\bibfield  {journal} {\bibinfo  {journal}
  {Phys. Rev. D}\ }\textbf {\bibinfo {volume} {108}},\ \bibinfo {pages}
  {036020} (\bibinfo {year} {2023})},\ \Eprint
  {http://arxiv.org/abs/2305.18677} {arXiv:2305.18677 [hep-ph]} \BibitemShut
  {NoStop}%
\bibitem [{\citenamefont {Cyncynates}\ and\ \citenamefont
  {Thompson}(2023)}]{Cyncynates:2023esj}%
  \BibitemOpen
  \bibfield  {author} {\bibinfo {author} {\bibfnamefont {David}\ \bibnamefont
  {Cyncynates}}\ and\ \bibinfo {author} {\bibfnamefont {Jedidiah~O.}\
  \bibnamefont {Thompson}},\ }\bibfield  {title} {\enquote {\bibinfo {title}
  {{Heavy QCD axion dark matter from avoided level crossing}},}\ }\href
  {\doibase 10.1103/PhysRevD.108.L091703} {\bibfield  {journal} {\bibinfo
  {journal} {Phys. Rev. D}\ }\textbf {\bibinfo {volume} {108}},\ \bibinfo
  {pages} {L091703} (\bibinfo {year} {2023})},\ \Eprint
  {http://arxiv.org/abs/2306.04678} {arXiv:2306.04678 [hep-ph]} \BibitemShut
  {NoStop}%
\bibitem [{\citenamefont {Li}(2023)}]{Li:2023xkn}%
  \BibitemOpen
  \bibfield  {author} {\bibinfo {author} {\bibfnamefont {Hai-Jun}\ \bibnamefont
  {Li}},\ }\bibfield  {title} {\enquote {\bibinfo {title} {{Axion dark matter
  with explicit Peccei-Quinn symmetry breaking in the axiverse}},}\ }\href@noop
  {} {\  (\bibinfo {year} {2023})},\ \Eprint {http://arxiv.org/abs/2307.09245}
  {arXiv:2307.09245 [hep-ph]} \BibitemShut {NoStop}%
\end{thebibliography}%
\end{document}